\documentclass[aps,preprint,nofootinbib]{revtex4}
\usepackage{epsfig}

\def\siml{{\ \lower-1.2pt\vbox{\hbox{\rlap{$<$}\lower6pt\vbox{\hbox{$\sim$}}}}\ }}
\def\simg{{\ \lower-1.2pt\vbox{\hbox{\rlap{$>$}\lower6pt\vbox{\hbox{$\sim$}}}}\ }}

\def\vbfD{{\ \lower-8pt\vbox{\hbox{\rlap{$\!\leftrightarrow$}\lower8pt\vbox{\hbox{$\!\bf D$}}}}\ }}
\def\dsl{\,\raise.15ex\hbox{/}\mkern-13.5mu D}

\newcommand{\nn}{\nonumber}
\newcommand{\be}{\begin{equation}}
\newcommand{\ee}{\end{equation}}
\newcommand{\bea}{\begin{eqnarray}}
\newcommand{\eea}{\end{eqnarray}}
\newcommand{\beq}{\begin{equation}}
\newcommand{\eeq}{\end{equation}}
\newcommand{\bqa}{\begin{eqnarray}}
\newcommand{\eqa}{\end{eqnarray}}

\newcommand{\Appendix}[1]%
    {%
     \section{#1}%
      }

\begin{document}
\preprint{\vbox{\halign{ &# \hfil\cr &\today\cr }}}
\title{Variational Study of Weakly Coupled Triply Heavy Baryons}
\author{Yu Jia}
\affiliation{Institute of High Energy Physics, 
Chinese Academy of Sciences, Beijing 100049, China}

\begin{abstract}
Baryons made of three heavy quarks become weakly coupled, when
all the quarks are sufficiently heavy such that 
the typical momentum transfer
is much larger than $\Lambda_{\rm QCD}$. 
We use variational method to estimate  masses of 
the lowest-lying
$bcc$, $ccc$, $bbb$ and $bbc$ states by assuming 
they are Coulomb bound states.
Our predictions for these states 
are systematically
lower than those made long ago by Bjorken.
\end{abstract}


\maketitle

\newpage

\section{Introduction}

The field of heavy quark spectroscopy is experiencing a 
rapid renaissance, mainly propelled by the emergencies of 
several unusual charmonium resonances,
of which $X(3872)$, $Y(4260)$ are the highlights~\cite{Swanson:2006st}.
Accompanied with these unexpected discoveries, progress has 
also been made steadily in the more traditional sector of 
charmonium spectroscopy, exemplified by the
recent sightings of several long-awaited particles 
such as $\eta_c(2S)$, $h_c$,
and particularly the doubly-charmed baryons such as 
$\Xi_{cc}^+$, $\Xi^{++}_{cc}$. 
Precise knowledge of their properties will help to 
refine our present understanding of 
heavy quark dynamics~\cite{Brambilla:2004wf}.

After the tentative establishment of 
the doubly charmed baryons~\cite{Moinester:2002uw}, 
one may naturally expects to 
fill the baryon family with the last missing member, {\it i.e.}, 
baryons composed entirely of heavy quarks,  
denoted the $QQQ$ states in short.
Being a baryonic analogue of heavy quarkonium, 
the triply-heavy baryons are of considerable theoretical interest, 
since they are free of light quark contamination 
and may serve as a clean probe to the interplay between 
perturbative and nonperturbative QCD.

One of the basic properties of these heaviest baryons in Nature
is their masses, which will be the primary concern of this paper. 
In contrast with the spectra of the singly-heavy and 
doubly-heavy baryons, to which a vast number of literature 
based on either phenomenological 
approaches or lattice QCD simulations are dedicated,  
only sparse attention has been paid to the 
spectroscopy of triply-heavy baryons, perhaps
mainly due to the lack of experimental incentive.

The interest toward these baryons 
can be traced back to Bjorken, 
who first carried out a comprehensive studies 
on their  properties two decades ago,
particularly focusing on the discovery 
potential of the  triply-charmed baryon state~\cite{Bjorken:1985ei}.
Reconstructing a $QQQ$ candidate is a rather challenging 
job experimentally, since it is difficult to 
separate all the decay products emerging from the 
cascade decay chain  $QQQ\to QQq \to Qqq$ 
from the copious hadronic background. 
Nevertheless, according to Bjorken, some semileptonic 
decay channels
such as $\Omega_{ccc}^{++}\to \Omega^- + 3\mu^++3\nu_\mu$, 
may offer a clean signature for a $ccc$ event.

Needless to say, the discovery potential of triply-heavy baryons
also crucially depends on the production environment. 
Baranov and Slad have shown 
that the production cross sections for triply-charmed
baryons at $e^+ e^-$ collider are too tiny 
to be practically relevant~\cite{Baranov:2004er}.
Gomshi-Nobary and Sepahvand have 
recently calculated the fragmentation functions of $c$ and $b$ 
evolving into various triply-heavy baryons, 
and estimated that  the corresponding fragmentation probabilities 
vary in the range $10^{-7}\sim 10^{-4}$~\cite{GomshiNobary:2004mq}.  
They consequently estimated two largest cross sections, 
which are associated with producing 
$\Omega_{bcc}$ and $\Omega_{ccc}$, 
to be about 2 and $0.3$ nb
in the forthcoming Large Hadron Collider (LHC) experiment with 
cuts of $p_T>10\;{\rm GeV}$ and $|y|<1$. 
For an integrated luminosity of 300 ${\rm fb}^{-1}$ 
(about one year of running at 
the LHC design luminosity ${\cal L}=10^{34} \,{\rm cm}^2 \,{\rm s}^{-1}$), 
the amount of
$\Omega_{bcc}$ and $\Omega_{ccc}$ yield can reach about 
$6\times 10^8$  and $1\times 10^8$.
It seems rather promising to establish these two states 
in such a large data sample.

Stimulated by the discovery possibility of triply-heavy baryons 
in near future, it is no longer of only academic interest 
to study their properties like mass spectra.
Unfortunately, no predictions to the masses of triply 
heavy baryons from lattice QCD 
simulations have emerged yet (only the static three-quark 
potential has been measured~\cite{Bali:2000gf,Takahashi:2000te}), 
and one has to resort to other theoretical means
at this moment.

Heavy quarkonium spectroscopy is traditionally the arena 
of phenomenological potential models, 
which in general incorporate a long-range 
confinement 
interaction~\cite{Eichten:1978tg,Richardson:1978bt,Buchmuller:1980su}
(see also Godfrey's contribution in \cite{Brambilla:2004wf}).
Nevertheless, recent advances in 
nonrelativistic effective field theories of QCD,
particularly the effective theory dubbed 
potential NRQCD (pNRQCD),  has started to put 
heavy quarkonium spectroscopy on a 
model-independent ground~\cite{Pineda:1997bj}
(for a recent review, see \cite{Brambilla:2004jw}).
A novel aspect of this effective field 
theory is that the interquark potential 
arises as the matching coefficients.   
In this language, different quarkonium states 
are categorized with respect to the 
relative magnitude between 
the typical momentum scale, $mv$,  and 
the nonperturbative QCD
scale, $\Lambda_{\rm QCD}$.
In the case of $mv\gg \Lambda_{\rm QCD}$, the corresponding 
state is said to be {\it weakly coupled}, 
and the dynamics is largely dictated by the short-distance
potential which can be calculated order by order in $\alpha_s$;
in the other situation  like $mv \sim \Lambda_{\rm QCD}$,
the state is said to be {\it strongly coupled}
since the potential is no longer calculable in 
perturbation theory,
instead must be determined by nonperturbative methods 
such as lattice QCD.
It is in the latter situation 
that a pNRQCD framework intimately resembles
the phenomenological Cornell model.
Evidences are accumulating to hint that $\Upsilon$, $B_c$ may  well 
be identified with the weakly-coupled system,
whereas $J/\psi$ lies in the borderline between 
the weak and strong couping regime, and the first 
few excited bottomonium 
states (far from the open flavor threshold) 
belong to the strongly-coupled system~\cite{Brambilla:2004jw}.

In parallel with the formulation of pNRQCD for the quarkonium, 
Brambilla,  Vairo and Rosch 
have recently laid down an analogous framework
for triply-heavy baryons~\cite{Brambilla:2005yk}.
The effective Lagrangian has been written down   
for both weakly-coupled and strongly-coupled 
$QQQ$ states, with some of 
the matching coefficients supplied.
Among various possible applications of Ref.~\cite{Brambilla:2005yk}, 
exploring the mass spectra of $QQQ$ states is the 
most straightforward to think of.
This is the very goal of the present work. 
To make things more tractable, we will confine ourselves 
in this work to the weakly-coupled states only. 
The $\Omega_{ttt}$, if exists, would be an ideal prototype for
such a state. However, to be phenomenologically relevant,
we have to stick to  baryons 
made exclusively of bottom or charm. 
As in quarkonium, most probably only the ground states
are amenable to a weak-coupling assignment.
To be objective, due to  weaker interquark 
color strength in a baryon than in a meson,
and not so heavy charm and bottom masses, 
one cannot exclude the possibility that even 
the ground states might be strongly coupled.

Despite this disclaimer, 
we will proceed by assuming that the $QQQ$ ground states
are indeed the weakly-coupled system. 
In this work, we attempt to estimate the 
leading order contribution to the binding energy,  
therefore for this purpose,
only the tree-level static inter-quark potential, {\it i.e.},
Coulomb potential, needs to be considered.
Since rigorously solving a three-body 
Coulomb bound state problem is beyond
our current ability, we have to resort to 
some sort of approximation method.
Stimulated by  success of the 
variational method in
coping with few-body atomic system, 
we will invoke this simple but efficient 
approximation scheme 
to address our baryonic problem. 
For baryons containing simultaneously $b$ and $c$,  
we will take advantage of the mass hierarchy 
$m_b\gg m_c$ to guide our variational analysis,
just in analogy with that in the simple 3-body
atomic system such as helium atom
and the ionized hydrogen molecule, 
the physical picture becomes
much more tractable by exploiting the fact
$m_N\gg m_e$.

The rest of the paper is distributed as follows. 
In Section~\ref{WC:QQQ}, 
we present a brief introduction to the most
relevant features of 
the triply-heavy baryons in the weak-coupling regime.  
In Section~\ref{VC:MainBody},  which is the main body of
this work, we perform a detailed variational analysis 
to the binding energy of various $QQQ$ ground states.  
Three different classes of triply-heavy baryons, 
$bcc$, $ccc$ ($bbb$) and $bbc$ states are treated 
separately,  with the hierarchy  
$m_b\gg m_c$ utilized as a guidance 
for choosing proper trial states.
Considerable amount of effort 
has been devoted to the $bbc$ state,
which is the most interesting case, but
also most difficult to analyze.
We have employed three different approaches 
to study this state and
explored  the implication of each approach 
in depth. Particularly the relevance of
the compact diquark picture 
is discussed.
In Section~\ref{Phenomenology}, 
we present our predictions to the masses of 
all the lowest-lying triply heavy baryons, 
and compare our results with other work. 
We summarize and present an outlook 
in Section~\ref{Summary}.

\section{Weakly coupled $QQQ$ states}
\label{WC:QQQ}

In this section we recapitulate the major aspects of
weakly-coupled triply heavy baryons which are most relevant
to this work. For more comprehensive discussion 
from the perspective of pNRQCD,
we refer the interested readers to 
Ref.~\cite{Brambilla:2005yk,Brambilla:2004jw}.

To efficiently investigate the low energy properties
of a tripled-heavy baryon, such as binding energy, 
it is convenient to work with a 
low energy effective theory
that focuses on the most relevant degrees of freedom.
In a weakly-coupled $QQQ$ state, the relevant
low energy degrees of freedom are  
nonrelativistic heavy quarks and (ultrasoft) gluons 
with energy and momentum of order $m v^2$, just 
like in a weakly-coupled quarkonium.
All the high energy degrees of freedom, 
which can only appear in virtual states, 
have been integrated out explicitly. 
One particularly important high energy mode is
the (soft) gluons with momentum of order $mv$,
whose effects are encoded in the low energy theory
as the interquark potentials.
Since we have $mv \gg \Lambda_{\rm QCD}$
in a weakly coupled state, the 
potentials can be determined  in perturbation theory 
by matching procedure.
There are infinite number of potentials, 
which are organized 
in  expansions of  $1/m$.
The most important potential is the 
${\cal O} (1/m^0)$
(static) potential.

The explicit form of potentials depends on the overall 
color configuration of quarks. 
Three heavy quarks can be in  
either  color singlet, octet or decuplet state.
The color-singlet state represents the  
most important case,  
since it constitutes the leading Fock component 
of a baryon. 
The singlet static potential is well known,
\bqa
 V_S^{(0)} &=& 
- {2 \alpha_s\over 3} \left( {1\over |{\bf x}_1- {\bf x}_2|} 
+{1\over |{\bf x}_2- {\bf x}_3|}+ 
{1\over |{\bf x}_3- {\bf x}_1|}\right)+ {\cal O}(\alpha_s^2)\,,
\eqa
where the color interaction between any pair of quarks  
is attractive.
In general,  in the color octet and decuplet
configurations, some or all pairs of quarks will
repel each other.

Thus far, the low energy effective theory is completely 
depicted by a set of uncoupled Schr\"{o}dinger equations
governing the motion of heavy quarks 
in different color configurations.
The situation becomes more intriguing when 
ultrasoft gluons are included. 
Since the typical wavelength of ultrasoft gluons
is much longer than the typical interquark distance,
the gluon fields can be multipole expanded.
Very much like the electromagnetic multipole transition
in atoms,  ultrasoft gluons can also induce 
chromo-electromagnetic multipole transition
from one heavy quark color configuration to a different one.  
In particular, a chromo-electric dipole transition 
can occur between a color-singlet $QQQ$ configuration 
to an octet one.
The interaction between bound heavy quarks and 
vacuum gluonic fluctuations 
through this chrome-$E1$ operator,
will  generate the leading nonperturbative correction 
to mass of a triply-heavy baryon,
as  a manifestation of 
Lamb shift in QCD~\footnote{The analogous effect
in quarkonium system, 
originally considered by Voloshin~\cite{Voloshin:1978hc},
has been extensively explored by many authors.}.
The magnitude of this nonperturbative correction
depends on the relative size between $mv^2$ and 
$\Lambda_{\rm QCD}$. 
It is quite difficult to estimate this effect accurately, 
and we will not consider it further.

We end this section by commenting briefly on the 
solidity of the weak-coupling assignment to the
lowest-lying triply heavy baryons
that are of practical interest.
As was admitted in Introduction, 
since the interquark color strength in a baryon  
is only a half of the quark-antiquark color strength
in a quarkonium, the typical dimension of a 
triply heavy baryon, say, $\Omega_{bbb}$, 
is expected to be considerably  fatter than that of $\Upsilon$. 
One may worry that the wave function of the 
former could
 permeate deeply into the confinement region. 
Fortunately, as will be shown quantitatively 
in the forthcoming section,
the interquark attraction is effectively enhanced 
due to the influence of the third quark,
so the actual situation turns out to be
considerably better than 
this pessimistic anticipation.

\section{Variational Estimate of Binding Energy}
\label{VC:MainBody}

In this section, we attempt to estimate the binding
energy of various $QQQ$ ground states.
Our starting point is the color-singlet 
hamiltonian:
\bqa
H_S &=& -{1\over 2} \,\sum_{i=1}^3 {\nabla_i^2\over m_i} 
+ V_S^{(0)}+ \cdots\,,
\label{H:3-body:general}
\eqa
where the ellipsis stands for the higher-dimensional potentials 
suppressed by powers of $1/m$. 
Because the purpose of this work is 
to calculate the leading ${\cal O}(\alpha_s^2)$
contribution, 
we will restrict to 
the lowest order static potential only.

To describe a bound state, we need first separate 
the relative motions of quarks  from  
the center-of-mass motion  in (\ref{H:3-body:general}).
There are infinite ways to 
perform this separation.
A simple way is to replace the old coordinates 
by the center-of-mass coordinate plus 
two new coordinates defined
as the positions of the quark 1, 2 
relative to the quark 3:
\bqa
{\bf X}& =&  {1\over \sum m_i}\,\sum_{i=1}^3 m_i {\bf x}_i \,,
\nn \\
{\bf r}_1 &=& {\bf x}_1- {\bf x}_3\,,
\nn \\
{\bf r}_2 &=& {\bf x}_2- {\bf x}_3\,.
\label{bcc:coordinate}
\eqa
In terms of these new coordinates, the  
hamiltonian (\ref{H:3-body:general}) can be separated into
\beq
     H_S =H_S^{\rm CM} + h_S\,,
\eeq
where $H_S^{\rm CM}=-\nabla^2_{\bf X}/ (2\sum m_i)$ 
is the center-of-mass part,  and the part governing 
the relative motion reads
\bqa
h_S &=& -{\nabla_{r_1}^2\over 2 m_{13}} -
{\nabla_{r_2}^2\over 2 m_{23}} - 
{\nabla_{r_1} \cdot \nabla_{r_2} \over m_3} 
- {2\alpha_s\over 3}
\left({1\over r_{1}}+ {1\over r_{2}}+{1\over r_{12}}\right)\,,
\label{Hr:three:body}
\eqa
where $r_{12}=|{\bf r}_1-{\bf r}_2|$, $m_{ij}=(1/m_i+1/m_j)^{-1}$ 
is the reduced mass between quark
$i$ and $j$. 
In such a coordinate system, the quark 3,
sitting at the origin,
is artificially singled out
from two other quarks.

Our task then becomes solving the 
bound state problem
defined in (\ref{Hr:three:body}).
In the following, we will use the
variational method to estimate the 
corresponding binding energy 
of each type of 
$QQQ$ ground states.

\subsection{$bcc$ }

\begin{figure}[bt]
  \centerline{\epsfysize= 5.5truecm \epsfbox{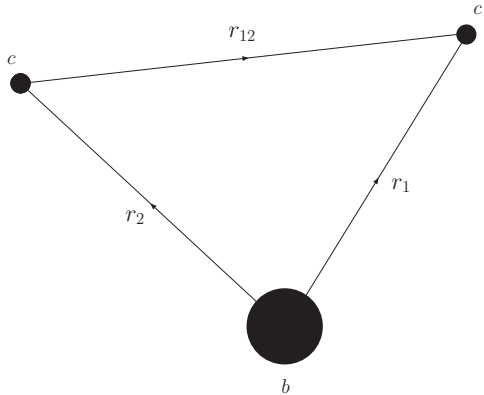}  }
 {\tighten
\caption{
Sketch of the coordinate system used for the $bcc$ state.
 }
\label{sketch:bcc} }
\end{figure}

We start from the simplest case,  the baryon  
made of one heavier bottom quark of mass $M$
and  two lighter charm quarks of mass $m$
(throughout this work, we will use the notation 
$M\equiv m_b$ and $m\equiv m_c$).

It is convenient to choose a coordinate system 
as specified in (\ref{bcc:coordinate}), 
with $b$ sitting at the origin.
This coordinate system is sketched in Fig.~\ref{sketch:bcc}. 
Subsequently, substituting $m_1=m_2=m$ and $m_3=M$ into
(\ref{Hr:three:body}),  the hamiltonian describing 
the internal motion  of a singlet $bbc$ state reads
\bqa
h_S &=& -{\nabla_1^2\over 2 m_{\rm red}} -
{\nabla_2^2\over 2 m_{\rm red}} - 
{\nabla_1 \cdot \nabla_2 \over M} 
- {2\alpha_s\over 3}
\left({1\over r_1}+{1\over r_2}\right)-{2\alpha_s\over 3}
{1\over r_{12}}\,,
\label{Hamilt:bcc}
\eqa
where $m_{\rm red}=(1/m +1/M)^{-1}$ is the reduced mass between 
$c$ and $b$. Note in this choice of coordinates,
the motion of $b$ is embodied in the reduced mass
and the operator $\nabla_1 \cdot \nabla_2 / M$.

Let us first consider an ideal $bcc$ state 
with $M/m\to \infty$. In this situation, 
the $b$ quark just acts 
as a static color source,
with two $c$ quarks revolving around.
This picture is very similar
to that of the two-electron atoms such as 
$H^-$, $He$ and $Li^+$,
where the nucleus is practically 
fixed in space, and two K-shell electrons 
orbit about it.
Estimating the energy of the two-electron
atoms is considered as a classical 
application of the variational method,  
which has been discussed  
virtually in every 
quantum mechanics textbook ({\it e.g.}, see~\cite{Baym:QM}).

Closely following the textbook treatment 
of helium,  
we may approximate  the $bcc$ ground state 
to be the one in which each of the 
$c$  moves in the $1s$ orbital 
of an effective Coulomb potential, 
somewhat stronger than
$- 2\alpha_s/ 3r$.
This is so because the attractive
color interaction felt by each $c$ due to $b$
is strengthened due to the 
attraction exerted by another $c$. 
This is in opposite situation 
to $He$, where  nuclear charge felt by each electron 
is partly screened due to the repulsion
exerted by another electron.

In a physical $bcc$ state, the hierarchy $M \gg m$ 
is much less perfect than that in a helium.
Nevertheless, the above ansatz about the 
form of the ground state wave function
still seems plausible.
What we need is  to take 
the motion of $b$ into account. 
For notational simplicity, we will take
the ``baryonic" unit  $m_{\rm red}= 2 \alpha_s/3=1$,
in which all the length and energy scales are 
measured in the unit of the Bohr radius 
$(m_{\rm red} \,2\alpha_s/ 3 )^{-1}$
and Bohr energy $m_{\rm red} \,(2\alpha_s/ 3)^2$.
We choose the spatial part of 
trial wave functions as 
\beq
    \Psi({\bf r}_1, {\bf r}_2) = f(r_1) f(r_2) \,,
\label{Psi:bcc}
\eeq
where 
\beq
    f(r) = {\lambda^{3/2}\over \sqrt{\pi}} e^{-\lambda r} 
\label{Coulomb:1s:wvf}    
\eeq
is the normalized $1s$ Coulomb wave function. 
Here $\lambda$ is a variational parameter, which 
characterizes the effective color charge 
of $b$ perceived by each of the $c$.
Obviously, when the attractive interaction between two charm 
is turned off, $\lambda$ would simply be 1.
For the reason discussed earlier, 
we expect $\lambda>1$ in our case, 
so that each $c$ can be thought of
moving on a squeezed $1s$ orbital. 
This is opposite to what is expected in a helium.

In the $He$ ground state, 
two K-shell electrons must form a spin singlet
to obey Fermi statistics, 
since its spatial wave function is symmetric
under the interchange of two electrons. 
Due to the extra color degree of freedom carried by quarks, 
two $c$ quarks in the $bcc$ ground state  
must instead be a spin triplet.
When combined with $b$, the lowest-lying $bcc$ baryon
can be either $J^P={1\over 2}^+$ or $J={3 \over 2}^+$, 
which are degenerate up to ${\cal O}(m^2 \alpha_s^4/M)$ 
corrections due to the hyperfine splitting.

We now attempt to find the expression 
for 
the ground state energy, $E$.
Taking the expectation value of $h_S$ with
the trial wave function in (\ref{Psi:bcc}),
after some effort one obtains
\beq
    E =  -\lambda^2 + 2\lambda (\lambda-1) + {\cal J}\,,
\label{Ebin:bcc}
\eeq
where 
\beq
   {\cal J} = -{\lambda^6\over \pi^2} \int\!\!\!\int d^3 r_1 d^3 r_2 \,
   {e^{-2\lambda (r_1+r_2)}\over r_{12}} = -{5\over 8}\lambda\,,
\label{J:definition}
\eeq
measures the average potential energy
stored between two charm quarks.

Note that the double integral involving the 
$\nabla_1 \cdot \nabla_2$ term
vanishes, because of spherical symmetry 
possessed by the $1s$ wave functions.
Thus,  the effect of kinetic energy of $b$ is 
fully taken into account 
by the reduced mass $m_{\rm red}$.

The minimum of energy can be found 
by requiring $d E /d\lambda=0$, 
which leads to
\bqa
     \lambda &=&  {21\over 16}\,,
\label{lmd:bcc:variation:determ}      
\eqa
indeed compatible with our expectation.
The corresponding ground state energy is
\bqa
     E &=&  - \left({21\over 16}\right)^2 
     \longrightarrow 
     - \left({7\over 8}\right)^2 
     m_{\rm red} \,\alpha_s^2\,,
\label{E:bcc:variation:determ}
\eqa
where the Bohr energy has been inserted 
in the last term,
to recover the actual dimension of energy.

\subsection{$ccc$}

The triply charmed baryon states  no longer 
have an atomic counterpart. 
On the other hand, the $ccc$ ground state is 
highly constrained by symmetry.
To have lowest energy, it necessarily
possesses  a totally symmetric 
spatial wave function. 
After the totally antisymmetric 
color wave function is included, 
Fermi statistics then demands that it
must have $J^P={3\over 2}^+$.

We again work with a coordinate system  
defined in (\ref{bcc:coordinate}), with
one $c$, 
artificially denoted charm 3, 
fixed at the origin. 
The hamiltonian depicting the relative motion of
three identical $c$ can be obtained by 
making the replacement  $M\to m$ 
in (\ref{Hamilt:bcc}).
We then have the reduced mass $m_{\rm red}=m/2$.
To condense the notation, we will work with
the ``baryonic" unit, in which 
the corresponding hamiltonian becomes
\bqa
h_S &=& -{\nabla_1^2\over 2} -
{\nabla_2^2\over 2} - 
{\nabla_1 \cdot \nabla_2 \over 2} 
- \left({1\over r_1}+{1\over r_2}+{1\over r_{12}}\right)\,.
\label{Hamilt:ccc}
\eqa

We are attempting to seek a proper form for  
the trial wave function for $ccc$ ground state.
One simplest choice is  
motivated from that adopted 
for a $bcc$ state.
Let us temporarily imagine the charm 3 
can be distinguished from the rest of two, 
then (\ref{Psi:bcc}) constitutes a reasonable 
representation for such a state.
Now coming back to a physical $ccc$ state,
to account for the indistinguishablity
of $c$, 
we should fully symmetrize (\ref{Psi:bcc}).
With the spin part of 
wave function suppressed, 
the trial wave function then reads
\beq
    \Psi({\bf r}_1, {\bf r}_2) = {f(r_1) f(r_2)+ 
    f(r_1) f(r_{12})+f(r_{12}) f(r_2)
    \over \sqrt{3\,(1+2\,{\cal T})} } \,,
\label{Psi:ccc}    
\eeq
where $f$ is the same as given in (\ref{Coulomb:1s:wvf}), 
and contains a variational parameter $\lambda$.    
One can check this wave function is 
symmetrical under the interchange 
of any two charm quarks.  
The $\Psi$  is normalized by incorporating
the overlap integral $\cal T$,
\bqa  
  {\cal T} &=& {\lambda^6\over \pi^2}
  \int\!\!\!\int\!d^3 r_1 d^3 r_2 \,
   e^{-\lambda (2r_1+r_2+r_{12})} = {176\over 243}\,.
\eqa

Taking the expectation value of $h_S$ 
in the trial state (\ref{Psi:ccc}),
after some straightforward manipulation, 
we end up with the expression 
\bqa
  E &=& { -\lambda^2 + 2\lambda (\lambda-1) + {\cal J}
  -\lambda^2 {\cal T} + 2(\lambda-1) {\cal P}
  -4 {\cal Q}+{\cal F}-{\cal G}
  \over 1+ 2\,{\cal T} }\,,
\label{Ebin:ccc} 
\eqa
where ${\cal J}$ has been given in (\ref{J:definition}), 
and
\bqa
  {\cal P} &=& {\lambda^6\over \pi^2} \int\!\!\!\int\! d^3 r_1 d^3 r_2 \,
   {e^{-\lambda (2r_1+r_2+r_{12})}\over r_1 } = {68\over 81}\lambda \,,
    \nn  \\
   {\cal Q} &=& {\lambda^6\over \pi^2} \int\!\!\!\int\!  d^3 r_1 d^3 r_2  \,
   {e^{-\lambda (2r_1+r_2+r_{12})}\over r_2 } 
   \nn \\
   &= & 
  {\lambda^6\over \pi^2} \int\!\!\!\int\! d^3 r_1 d^3 r_2 \,
   {e^{-\lambda (2r_1+r_2+r_{12})}\over r_{12} } = {16\over 27}\lambda \,.  
   \nn \\
  {\cal F} &=& {\lambda^6\over \pi^2} \int\!\!\!\int\! d^3 r_1 d^3 r_2 \,
   e^{-\lambda (2r_1+r_2)} \,\nabla_2^2 \, e^{-\lambda r_{12}} 
   =- {112\over 243} \lambda^2 \,,
  \nn \\ 
  {\cal G} &=& {\lambda^6\over \pi^2} \int\!\!\!\int\!  d^3 r_1 d^3 r_2  \,
   e^{-\lambda (r_1+r_2)} \, \nabla_1 \cdot \nabla_2 \, e^{-\lambda (r_1+ r_{12})} 
   = {56\over 243} \lambda^2 \,.  
\eqa
Note the exchange integrals  
${\cal P}$, ${\cal Q}$ ${\cal F}$ and ${\cal G}$
arise from the symmetrization effect,  
which are absent in the expression for 
the energy of the $bcc$ baryon, (\ref{Ebin:bcc}). 
In particular, ${\cal G}$, the double integral 
involving $\nabla_1 \cdot \nabla_2$, 
no longer vanishes this time.

Substituting the results of 
these integrals into (\ref{Ebin:ccc}),
we obtain
\beq
   E = -{2\,595\over 952} \lambda + {531\over 595} \lambda^2 \,.
\eeq

The optimum can be found by variational principle,
\bqa
    \lambda &=& {4\,325\over 2\,832} \approx 1.527\,,
\label{lmd:ccc:variation:determ}  
\eqa
and 
\bqa
    E &=& -{3\,741\,125\over 1\,797\,376} 
    \longrightarrow  - 0.925 \, m_{\rm red} \,\alpha_s^2 \,,
\label{E:ccc:variation:determ}    
\eqa
where the normal unit is recovered in the last entity.

It is interesting to compare the results we have
got for the $bcc$ and $ccc$ ground states.
First lowering the $b$ mass in a $bcc$ state 
down to $m$, we get a fictitious $ccc$ state 
with one $c$ distinguishable from the other two.  
Comparing (\ref{E:ccc:variation:determ}) and  
(\ref{E:bcc:variation:determ}),
we immediately find the symmetrization effect tends to
lower the energy. Moreover, by comparing 
(\ref{lmd:ccc:variation:determ}) and
(\ref{lmd:bcc:variation:determ}), we find
the symmetrization effect also tends to 
compress the bound state size more.

For actual $bcc$ and $ccc$ states, 
we find $E_{\Omega_{ccc}}>E_{\Omega_{bcc}}$ 
(note $m_{\rm red}$ in two cases are different),
which implies that charm quarks in $\Omega_{bcc}$  
are more tightly bound than in $\Omega_{ccc}$.  
This is consistent with the general expectation
that a bound state with constitutes of vastly disparate 
masses is more stable than that with equal-mass constitutes,
say, a hydrogen atom is more stable than a positronium.

\subsection{$bbc$}

\begin{figure}[bt]
  \centerline{\epsfysize= 4.5truecm \epsfbox{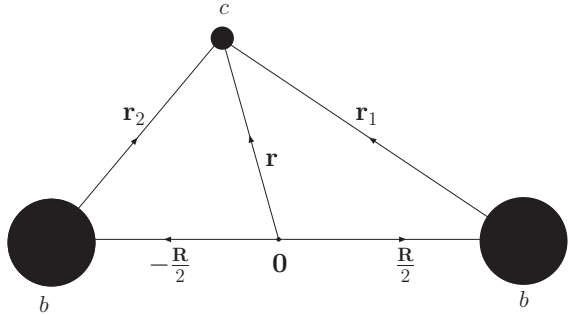}  }
 {\tighten
\caption{
Sketch of the coordinate system adopted for the $bbc$ state.}
\label{sketch:bbc} }
\end{figure}

We finally turn to  baryons made of two heavier $b$ quarks 
and a lighter $c$ quark. This type of baryon is  
more complicated than the preceding two, 
because the effective potential felt by $c$ 
is no longer 
spherically-symmetric,  but merely axially-symmetric.

To make the symmetry between two $b$ quarks manifest, 
we may adopt a more appropriate coordinate system 
other than (\ref{bcc:coordinate}). 
Letting $m_1=m_2=M$, $m_3=m$,  
we define the following new coordinates:
\bqa
{\bf X} &=& {M({\bf x}_1+ {\bf x}_2)+ m {\bf x_3} \over 2M+m}\,,
\nn \\
{\bf R} &=& {\bf x}_1- {\bf x}_2\,,
\nn \\
{\bf r} &=& {\bf x}_3- {{\bf x}_1 + {\bf x}_2\over 2} \,.
\label{New:Coord:bbc}
\eqa
Note now the coordinate origin  coincides 
with the middle point between two b quarks.
The geometry of these new coordinates can be clearly visualized 
in Fig.~\ref{sketch:bbc}.
Substituting  (\ref{New:Coord:bbc}) into the original hamiltonian 
(\ref{H:3-body:general}),
we find that the part of hamiltonian responsible for 
the internal motion is
\bqa
h_S &=& 
-{\nabla_R^2\over 2 M_{\rm red}} -{2 \alpha_s\over 3} {1\over R}
-{\nabla_r^2\over 2 m_{\rm red}} - {2 \alpha_s\over 3}
\left({1\over r_1}+{1\over r_2}\right)\,,
\label{Hamil:bbc:full}
\eqa
where $M_{\rm red}=M/2$, 
$m_{\rm red}= (1/ m+ 1/ 2M)^{-1}$ are the reduced masses, 
and $r_1=|{\bf r}- {{\bf R}\over 2}|$, $r_2=|{\bf r}+ {{\bf R}\over 2}|$.

A nuisance may deserve some caution 
before we move on further.
Two strong coupling constants  
in (\ref{Hamil:bbc:full})
have been tacitly assumed to be evaluated 
at the same renormalization scale $\mu$.
This procedure seems incompatible with 
our intuition that the first $\alpha_s$ 
should be affiliated with
a scale $\sim 1/\langle R \rangle$, and the second one
with a different scale $\sim 1/\langle r\rangle$, 
where $\langle R \rangle$ and $\langle r \rangle$ represent 
the typical values of $R$ and $r_i$, respectively.
When $M$ and $m$ are widely separated,
$\langle R \rangle\ll \langle r \rangle$ is expected,
and this recipe will miss the contributions
of large logarithm  
$\ln(\langle r \rangle/\langle R \rangle)$,
no matter which value of $\mu$ is chosen.
The symptom encountered in 
this equal-$\alpha_s$ ansatz
is a typical shortcoming of 
lowest-order perturbative calculation in a multi-scale 
problem, 
and one in principle can ameliorate its prediction
by appealing to renormalization group equation
to resum the large logarithms of the form 
$\alpha_s^n \ln^{n-1}(\langle r \rangle/\langle R \rangle)$.
Fortunately, for a physical $bbc$ baryon, 
$M$ is only three times larger than $m$, 
$\langle R \rangle$ and $\langle r \rangle$ 
likely don't differ much, 
hence we don't need worry much
about this nuisance.

In the following, we will treat 
the $bbc$ ground state with three different 
approaches: point-like diquark approximation, 
Born-Oppenheimer approximation, and variational
method.

\subsubsection{Point-like Diquark Approximation}
\label{Cmpact:Dq:Approx}

In a fictitious world where $M$ is 
many orders of magnitude heavier
than $m$, the physical
picture simplifies enormously.
The influence of $c$ to the motion of
very heavy $b$ can be safely neglected.
Consequently, two $b$ quarks in the $bbc$ ground state
form a $1s$ spin-triplet state.
The very compact Bohr radius of $b$, cannot be resolved 
by $c$ which is orbiting from far away. 
Thus from the perspective of $c$, 
the $bb$ cluster is just 
like a point particle.
To distribute itself in the lowest energy, 
the $c$ is revolving around this point particle 
in the corresponding $1s$ orbital.
Being in $\overline{3}$ color state, this compact diquark 
may be identified with a heavy antiquark.
In this sense, the $bbc$ ground state  is analogous to 
the heavy quarkonium $B_c$.

In passing, it is worth mentioning that 
the doubly heavy baryons, such as $bbq$ states,
fit into this compact diquark picture
to a better extent than the $bbc$ state,
because the average distance between $q$ and diquark in the former,
$\sim 1/\Lambda_{\rm QCD}$, 
is considerably larger than 
the average distance between $c$ and
diquark in the latter, $\sim 1/(m \alpha_s)$.
Properties of doubly-heavy baryons 
was first studied within a compact diquark picture 
long ago in HQET language~\cite{Savage:1990di}. 
Some refinement to this picture,   
which invokes the nonrelativistic EFT of QCD
to describe the internal excitation of the diquark, 
has recently 
come out~\cite{Brambilla:2005yk,Fleming:2005pd}.

The form of (\ref{Hamil:bbc:full}) is
particularly convenient to accommodate
the compact diquark picture~\footnote 
{For pedagogical purpose, 
in the following two $\alpha_s$ in (\ref{Hamil:bbc:full})
will be simply taken  
equal even in the limit $M/m\to \infty$.}.
Because $\langle R \rangle\ll \langle r \rangle$
in this case,
one may approximate $r_1$ and $r_2$ by $r$, 
the color potential felt by $c$
then becomes $- 4\alpha_s/3r$, 
as if it is due to a heavy antiquark sitting at the origin.
Eq.~(\ref{Hamil:bbc:full}) then collapses into 
two separate hamiltonians, one governing 
the internal motion of the diquark, the other governing
the motion of $c$ in a central Coulomb potential.  
The energy of the $bbc$ ground state is then
simply the sum
\bqa
E&=& -{1\over 2} M_{\rm red} \left({2\over 3} \alpha_s \right)^2-
{1\over 2} m_{\rm red} \left({4\over 3} \alpha_s \right)^2\,.
\label{compact:diquark:approx}
\eqa

For a physical $bbc$ state, 
the mass hierarchy between $b$ and $c$
is far from ideal,  so the usefulness of 
this oversimplified approximation
is doubtful.

\subsubsection{Born-Oppenheimer Approximation}

We now seek an alternative method that 
explicitly
incorporates the effect of finite diquark size.
First observe that an ideal $bbc$ state
bears some similarities with the simplest molecule, 
the $H_2^+$ ion, in the sense that both are
three-body bound states held together by Coulomb force, 
and both contain two heavy particles 
and one much lighter particle.
Motivated by this similarity, 
one may wonder whether some well-known method
developed to analyze $H_2^+$
can be transplanted here.

A standard tactics to cope with diatomic molecules, 
such as the $H_2^+$ ion,
is Born-Oppenheimer approximation 
({\it adiabatic} approximation). 
This method was originally motivated 
by the strong separation of 
time scales between
electronic and nuclear motion, 
which is mainly a consequence of 
the hierarchy $m_e \ll m_N$.
The recipe of this method 
is that, to solve molecular problem,
one first determines the electronic eigenstates 
at fixed nuclear positions, 
then takes the corresponding electronic energy 
as an effective potential, in conjunction with the 
internuclear Coulomb potential
to describe the nuclear motion.

There is a caveat, however. 
Despite the aforementioned similarities,
one should realize  
there is one fundamental difference between 
$H_2^+$ and the $bbc$ state,
that is,  the internuclear Coulomb interaction is repulsive,
whereas the Coulomb interaction between $b$ 
is attractive.
This difference in turn results in 
drastically distinct properties of $H^+_2$ 
and an ideal  $bbc$ state.
As a result, success of adiabatic approximation
to the former does not  automatically guarantee
that it can be taken for granted 
for the latter.

To better orientate ourselves, it is instructive 
to recall first how an adiabatic picture 
arises from the $H_2^+$ ground state~\cite{Baym:QM}.
A snapshot of this simplest molecule is that,
two nuclei slowly vibrate about 
some equilibrium positions with small amplitude, 
whereas the electron flies around 
much more swiftly. 
The vibrational nuclear motion  
is a consequence of the balance
between internuclear Coulomb repulsion and
an effective attractive interaction
induced by the electron.
A crucial fact is that the typical period 
of nuclear motion is much 
 ($\sim\sqrt{m_N/m_e}$)  {\it longer} than that of 
electronic motion.
It is thus a good approximation to regard
nuclei as frozen 
when considering the electronic motion,
consequently the electron will
distribute itself in the ground state
of this static nuclear potential.
Moreover, the electron can be regarded as
responding  instantaneously  to 
the change of nuclear arrangement,
therefore it  follows the nuclear motion 
{\it adiabatically},
which implies that it can always remain 
in the corresponding ground state for 
each nuclear configuration.

In contrast, an ideal $bbc$ state bears a completely 
different structure. As we have known, this state 
is characterized by a compact diquark picture.
The pull exerted by $c$  again induces an 
effective attractive interaction 
between two $b$ quarks.
However, when superimposed on the attractive 
Coulomb interaction, it helps,
though with a rather minor impact,
to push two $b$ closer.
The only agent to prevent a  
complete collapsing is the kinetic energy of $b$.
It is interesting to compare 
the overwhelmingly dominant role 
enjoyed by the kinetic energy of $b$
with the insignificant role played by the nuclear 
kinetic energy in $H_2^+$.

Based on the point-like diquark picture, 
one can show 
that $b$ is confined in a region about $m/M$ 
smaller than $c$,  the typical velocities of $b$
and  $c$ are about equal ($\sim \alpha_s$), 
and the typical kinetic energy of  $b$ 
is about  $M/m$  larger than that of $c$. 
Obviously, notions such as ``fast $c$" and 
``slow $b$" are simply 
misnomers.  Moreover, 
uncertainty principle tells that 
the typical orbiting period of $b$ is 
much ($\sim M/m$)  {\it shorter} than that of $c$.
As a result, $c$ can hardly follow the 
fuzzy pace of $b$, 
let alone to readjust itself instantaneously
to the ground state for a particular 
configuration of $b$.
In sharp contrast with $H_2^+$, the $bbc$ state 
exhibits a completely
{\it anti-adiabatic} nature.

The above negative argument seems to 
persuade us to 
give up adiabatic approximation
in analyzing an ideal $bbc$ state,
since the orthodox picture on which
this method is based is badly
violated.
Ironically, this method practically  
does yield correct result for this state.
The reason can be traced as follows.
We have argued that it is difficult for $c$ 
to react quickly to the rapid 
change of configurations of $b$. 
It simply gets confused.  
However, the really important point is, 
what $c$ can see is only a smeared $bb$ cluster
which is well localized in a small region,
it doesn't care about 
the details going on inside this cluster.
What $c$ can do is to distribute itself in
the ground state of the Coulomb potential due to a
remote $\overline 3$ color source.
This is of course nothing but the point-like 
diquark picture. 
Born-Oppenheimer method takes the energy eigenvalue
of $c$ in static configurations of $b$ 
as  effective potential for $b$. For an ideal $bbc$
state, only the value of this effective potential
at very small separation of $b$ is relevant,
which is just the $1s$ energy of $c$ in 
the Coulomb potential of an antiquark.
Following the Born-Oppenheimer procedure, 
the motion of $b$ is described by a new potential, 
which is the original Coulomb potential scaled up by
this tiny constant. 
One then readily reproduces the correct answer,
(\ref{compact:diquark:approx}),
for the ground state energy.

It is now clear that Born-Oppenheimer approximation 
practically works for an ideal $bbc$ state 
because of very compact diquark size.
But we certainly are more interested 
in the physical $bbc$ state.
Since $m$ and $M$ are not widely separated in this case,
there is no more strong separation of time scales,
this approximation thus is not 
expected to yield accurate result 
in the first place. 
Nevertheless, since this method takes the finite diquark size 
effect into consideration, which is relevant for a
physical $bbc$ state,
we will take a practical attitude, 
applying it to this state to 
watch what results will come out.

Let us now concretely analyze the $bbc$ state 
following Born-Oppenheimer method.
To start, we first approximate the 
full wave function  $\Psi$  as
\bqa
 \Psi({\bf R}, {\bf r}) &\approx &   \Phi ({\bf R})\, 
 \varphi( {\bf R},{\bf r}) \,,
\label{Expand:Full:Psi} 
\eqa
where $\varphi$ represents the charm ground state
for a static configuration of $b$, and $\Phi$
stands for the amplitude to find $b$ in
this configuration
when $c$ is in the state $\varphi$.

In the Born-Oppenheimer ansatz, 
we need first determine the lowest eigenstate 
$\varphi$ by solving the Schr\"{o}dinger equation
\bqa
\left[-{\nabla_r^2\over 2 m_{\rm red}} - {2 \alpha_s\over 3}
\left({1\over r_1}+{1\over r_2}\right)\right] \,  \varphi( {\bf R},{\bf r})
&=& \varepsilon(R) \,\varphi( {\bf R},{\bf r}) \,.
\label{Solution:Given:R}
\eqa
Since the positions of $b$ explicitly 
enter the potential, 
the energy eigenvalue $\varepsilon$  depends on $R$.

This is exactly the same problem as one 
encounters in $H_2^+$, to determine the 
electronic ground state 
at fixed nuclear positions,  
therefore we can follow the standard treatment~\cite{Flugge:QM}.
Solving (\ref{Solution:Given:R})
rigorously is unfeasible, one commonly appeals to
variational method. 
A reasonable form taken for the trial wave function 
is a linear combination of $1s$ charm states
centered on each of $b$ quarks.  
A variational parameter $\lambda$ is included
in the $1s$ trial state to characterize the 
effective color charge of $b$ perceived by $c$.  
Taking the indistinguishableness of $b$ 
into account, the trial wave function of $c$ 
takes the form
\bqa
  \varphi({\bf R}, {\bf r}) &=& {f(r_1) \pm f(r_2) \over
   \sqrt{2\,(1\pm {\cal S}(\lambda, R))}}\,,
\label{wv:c:bb:pm}   
\eqa
where $f$ is given in (\ref{Coulomb:1s:wvf}).
To keep our notation simple, we have chosen to 
work with the lighter ``baryonic" unit 
$m_{\rm red} = 2\alpha_s/ 3=1$.
The overlap integral is incorporated to 
make $\varphi$ normalized,
\bqa
  {\cal S}(\lambda, R) &=& {\lambda^3\over \pi} \int\!\! d^3 r \,
   e^{-\lambda \,(r_1+r_2)} = \left[1+\lambda R + 
   {\lambda^2 R^2 \over 3}\right]\, 
   e^{-\lambda R}\,.
\label{overlap:S}   
\eqa

\begin{figure}[bt]
  \centerline{\epsfysize= 9.0truecm \epsfbox{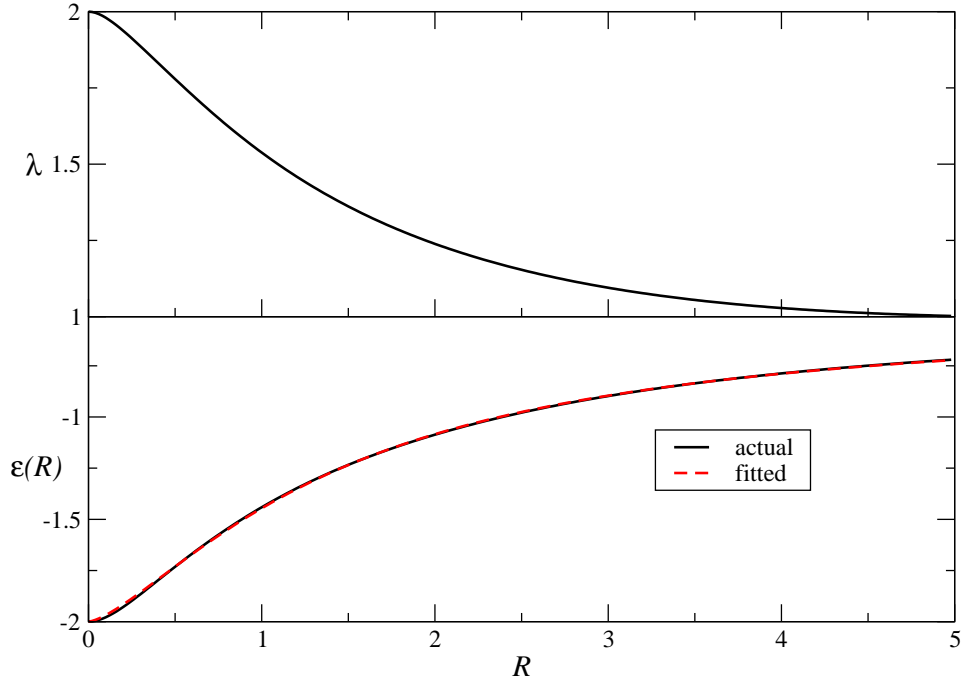}  }
 {\tighten
\caption{
$\lambda$ and effective potential 
as functions of $R$, 
determined by the variational calculus (solid line).
All the numbers are given in the 
lighter ``baryonic" unit.
In the lower half plot, 
the dashed curve represents the 
function given in 
(\ref{Effectivepotential:fitted}),
which is hardly distinguishable from the actual 
one.
}
\label{lmbda:Energy:R} }
\end{figure}

The wave function $\varphi$  must be either 
symmetric or antisymmetric
upon interchange of two $b$ (${\bf R}\to -{\bf R}$),
so that the corresponding $bb$ pair, 
if in relative $s$-wave,
must be either a spin triplet or singlet
in line with Fermi statistics. 
When combined with $c$, the former configuration 
corresponds to a $bbc$ state with 
$J^P={3\over 2}^+$ or ${1\over 2}^+$, and 
the latter corresponds to a state with 
$J^P={1\over 2}^-$.
As is well known in $H_2^+$, the antisymmetric configuration
has higher energy level than the symmetric one.
Since we are only interested in the 
$bbc$ ground state,  
we will discard the state with odd parity.

We thus choose the symmetric one
in (\ref{wv:c:bb:pm}).
Multiplying  both sides of (\ref{Solution:Given:R})
by the corresponding $\varphi^*$,  
integrating over $\bf r$,
we find that the charm energy reads
\bqa
\varepsilon(R) &=& -{\lambda^2 \over 2} +
{\lambda \,(\lambda-1)-{\cal C}(\lambda, R)  +
(\lambda-2)\, {\cal E}(\lambda, R)\over 1+ { \cal S(\lambda, R) }}\,.
\label{Eff:BO:bbc}
\eqa
The classical interaction integral ${\cal C}$
and exchange integral ${\cal E}$ are given by
\bqa
  {\cal C}(\lambda, R)  &=& {\lambda^3\over \pi} \int\!\! d^3 r \,
   {e^{-2 \, \lambda r_1} \over r_2} 
   = {1\over R}\left[ 1-(1+\lambda R)\,e^{-2\,\lambda R} \right]\,,
 \nn \\
  {\cal E} (\lambda, R) &=& {\lambda^3\over \pi} \int\!\! d^3 r \,
   {e^{-\lambda \,(r_1+r_2)} \over r_2} 
   = \lambda \, (1+\lambda R) \,e^{-\lambda R}\,.
\label{BA:C:E:integral}   
\eqa

To locate the minimum of (\ref{Eff:BO:bbc}) 
at a given $R$, we resort to the condition 
$\partial \varepsilon/\partial \lambda|_R=0$.
The analytical expression for the optimum, 
if can be worked out, would be very 
cumbersome, 
so we are content with providing 
numerical solutions only.
The optimized $\lambda$ and $\varepsilon$ 
as functions of $R$ are shown 
in Fig.~\ref{lmbda:Energy:R}.
In digression, we would like to mention that 
a trick  adopted by some texts 
(for example, \cite{Flugge:QM}),
which aims to facilitate finding the optimum, 
is mathematically inconsistent, therefore 
we have refrained from using it.

Fig.~\ref{lmbda:Energy:R} illustrates some 
expected features of charm ground state
in a static configuration of $b$.
At $R=0$, the $bb$ diquark shrinks to a point,
the color charges double, so 
we have $\lambda=2$  and $\varepsilon=-2^2/2=-2$.
This is exactly what we would expect by replacing a 
point-like diquark with an antiquark.
As $R$ gets large,  $c$ will be 
essentially localized with one of the $b$, 
forming a $1s$ state, and the influence 
of the other $b$ becomes negligible. 
To put in a quantitative way,
at large $R$, the effective charge
$\lambda \approx 1$, and the energy of $c$ 
is the $1s$ energy plus the potential energy between
$c$ and the other $b$, that is, 
$\varepsilon\approx -1/2-1/R$.

In the Born-Oppenheimer ansatz, 
the charm energy 
plays the role of effective potential for $b$.
To expedite our analysis,  
it is convenient to have an analytic formula
that mimics the actual $\varepsilon(R)$,
which is known only numerically. 
We find the following parameterization, 
\bqa
\varepsilon_{\rm fit}(R) &=& -0.5-
{1.5\over 1+ 0.586 \, R^{1.421}} \,,
\label{Effectivepotential:fitted}
\eqa
represents a good fit to the actual one,
with error less than one percent  
provided that $R$ is not too large.  
As already pointed out, 
due to the compact diquark 
nature of a $bbc$ state, 
only the knowledge 
in small $R$ range affects 
the bound state property.

The remaining task is to determine $\Phi$,
with the effective  potential 
taken as input.
In Born-Oppenheimer approximation, 
the motion of $b$ is simply governed by 
the following Schr\"{o}dinger equation\footnote{To 
arrive at this formula, one has dropped 
two additional terms containing $\nabla_R\, \varphi$ 
(see \cite{Baym:QM} for detailed derivation). 
This procedure is partly justified by the fact 
$\nabla_R \,\varphi\ll\nabla_R \,\Phi$
as expected from the compact diquark picture.
Although a rigorous mathematical proof is absent, 
we invoke the fact that this procedure makes 
correct prediction for an ideal $bbc$ state
as an evidence for its validity.
\label{fn:BA:approx:val}}:
\bqa
\left[-{\nabla_R^2\over 2 } - {1\over R} +\kappa\,
\left\{ -0.5-  1.5\,\left[1+ 0.586 \,
(\kappa\,R)^{1.421}\right]^{-1}
\right\}\right] \Phi({\bf R}) = E \, \Phi({\bf R}) \,,
\label{BA:approx:bbc:hvy:Bohrunit}
\eqa
where $\kappa\equiv m_{\rm red}/ M_{\rm red}$.
For convenience, we have switched to
the heavier ``baryonic" unit 
$M_{\rm red}= 2 \alpha_s/3=1$. 
Note $\kappa$ plays the role of scale conversion factor.

This equation can be solved numerically 
once $\kappa$ is specified. 
Consequently,  the energy of the 
baryon ground state, $E$,
can be identified with the 
eigenvalue of the corresponding $1s$ state.
The dependence of $E$ on $\kappa$ in a wide range
is shown in  Fig.~\ref{bbc:cmp:3method}.
As is expected, at small $\kappa$,
the energy predicted from this approach
does coincide with the one 
from the point-like diquark approximation.
Technically, this can be understood by
examining (\ref{BA:approx:bbc:hvy:Bohrunit})
in the limit $\kappa \to 0$,
in which the effective potential reduces to a 
constant $-2 \kappa$.
As discussed before, the underlying reason 
should be attributed to the fact 
that for small $\kappa$, only the value of 
the effective potential 
near $R=0$ is relevant.

In short, the lesson we have learned is that,
even though Born-Oppenheimer approximation 
is not theoretically justified for an ideal $bbc$ state, 
this procedure still leads to correct results   
because of 
the compact diquark nature in this state.

As $\kappa$ increases, the average diquark size
becomes comparable with 
the typical distance between $c$ and $b$.
In this situation, 
neither point-like diquark approximation
nor Born-Oppenheimer approximation is expected
to make reliable prediction.
Nevertheless, since the latter approach explicitly
incorporates the effect of finite diquark size, 
we may expect it is closer to the truth
than the former. 
As one can discern in Fig.~\ref{bbc:cmp:3method},
the prediction of $E$ from the latter approach
becomes incrementally higher than 
that from the former as $\kappa$ increases.
This is compatible with our expectation.
The larger $\kappa$ is, the more relevant 
the contribution of the effective potential
at large separation of $b$ becomes.
Since $\varepsilon$ monotonically increases with
$R$ (see Fig.~\ref{lmbda:Energy:R}), 
thus Born-Oppenheimer approximation 
predicts higher $E$.

\subsubsection{One-Step Variational Estimate}

We have shown that both the point-like diquark
approximation and Born-Oppenheimer approximation
render correct predictions  for an ideal $bbc$ state.
However, there is no {\it a priori} 
reason to expect them to work 
satisfactorily for a physical $bbc$ state,
where  $m$ and $M$ are not so widely separated.
A useful indicator is the ratio of the 
average diquark dimension to 
the typical distance between $c$ and $b$,
which is roughly 
\bqa
{\langle R \rangle \over \langle r \rangle } \sim 
 \kappa
\approx
{2 M_{J/\Psi} \over M_\Upsilon} \approx 0.6\,.
\eqa
Because of  poor separation between
$\langle R \rangle$ and $\langle r \rangle$,
a more general approach is
called for 
to analyze  the physical $bbc$ state.

In the following we will employ the third method, 
dubbed {\it one-step variational estimate}.
It takes basically the same 
variational ansatz 
as used for the $bcc$ and $ccc$ system.
However, due to more complex nature of the $bbc$ system,
two variational parameters 
have been introduced.
The term {\it one-step} implies that 
the ground state energy as well as
the full wave function are determined 
in a single step,
in contrast with 
Bohr-Oppenheimer procedure, in which 
one determines the  wave functions 
of $c$ and $b$ in two
successive steps.
On general ground, one expects 
this method is more
accurate than the other two, 
inasmuch as it is based entirely on 
the variational principle 
and no other approximation is invoked.
As long as the trial wave function is 
reasonably chosen, we expect 
it will render reliable prediction 
even when $\kappa$ is not small.

For notational convenience, we adopt the  
heavier  ``baryonic" unit  $M_{\rm red}= 2 \alpha_s/3=1$ here.
The hamiltonian (\ref{Hamil:bbc:full}) 
then simplifies to
\bqa
h_S &=& -{\nabla_R^2\over 2}  - {1\over R} 
-{1\over \kappa }\, \left({\nabla_r^2\over 2 } +
{\kappa\over r_1}+{\kappa\over r_2}\right)\,,
\label{bbc:onestep:hamiton:hvby}
\eqa
where the scale conversion factor is included
in the $c$ sector.

We first need to guess a proper 
form for the trial wave function $\Psi$.
It is natural to follow the 
ansatz of (\ref{Expand:Full:Psi}), 
to express $\Psi$ in a quasi-separable form  
$\Phi ({\bf R})\, \varphi({\bf R}, {\bf r})$, 
where $\Phi$ represents the $b$ wave function,
and $\varphi$ denotes the $c$
wave function, which may be taken  
the same as (\ref{wv:c:bb:pm}). 
This form of trial wave function clearly
embodies the point-like diquark picture
in the $\kappa\to 0$ limit.
Since $\varphi$ has incorporated 
the effects of finite diquark size, 
this choice of trial wave function 
seems reasonable also for large $\kappa$. 
We take the trial wave function 
for the $bbc$ ground state  explicitly to be 
\bqa
\Psi({\bf R}, {\bf r}) &=& {1\over \sqrt{2\,(1+\overline{\cal S})}}\,
{\delta^{3/2} \over \sqrt{\pi}}
{(\kappa\lambda)^{3/2} \over \sqrt{\pi}} \,e^{-\delta R}\,
\left(e^{-\kappa \lambda\,r_1}+e^{-\kappa \lambda\, r_2} \right)\,,
\label{bbc:trial:onestep:vr:est}
\eqa
with the spin wave function suppressed.
$\lambda$ and $\delta$ are 
variational parameters.
Note $\Psi$ is symmetric under 
the reflection ${\bf R}\to -{\bf R}$, 
as it should be for the ground state.
The wave function is normalized 
by incorporating the overlap integral 
\bqa
  \overline{\cal S} &=& {\delta^3\over \pi} \int\!\! d^3 R \,
   e^{-2\delta R} \, {\cal S}(\kappa \lambda, R) 
   = {16\,\delta^3\,
   (2\delta^2+5\delta\kappa \lambda +4\kappa^2\lambda^2)
   \over (2 \delta+\kappa\lambda)^5}  \,,
\eqa
where $\cal S$ is given in (\ref{overlap:S}).

The physical implication of $\lambda$
is the same as in Born-Oppenheimer ansatz,
which describes the effective charge of $b$ perceived
by $c$, except here it is taken as a
constant instead of a function of $R$.
This simplification seems plausible, at least for
small $\kappa$. 
As noticed before, the typical time scale 
characterizing  the change of  
configurations of $b$ is in general 
shorter than that of $c$, 
consequently $c$ only sees smeared trajectories of $b$.
When considering the impact of $b$ on $c$, 
it is reasonable to average its effects 
over different configurations of $b$.
This averaging procedure will lead to a 
constant value of $\lambda$.

The new parameter, $\delta$, 
is introduced simultaneously
to characterize the impact of $c$ 
on the geometry of the diquark.
It would simply equal 1 in the limit $\kappa\to 0$,
when the influence of $c$
becomes completely negligible.

\begin{figure}[bt]
  \centerline{\epsfysize= 9.0truecm \epsfbox{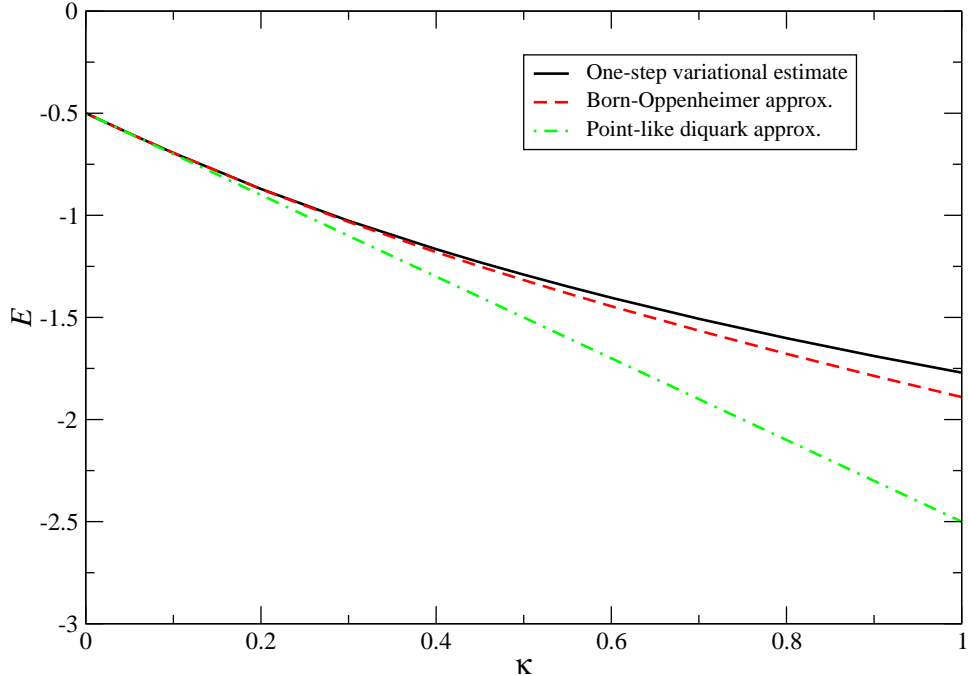}  }
 {\tighten
\caption{
The energy of the $bbc$ ground state (in the heavier 
``baryonic" unit)
as function of $m_{\rm red}/M_{\rm red}$.
Three curves are generated by implementing 
three different approximation
schemes. The dot-dashed line has the functional 
form $E=-{1\over 2}-2 \,\kappa$, 
as can be inferred from (\ref{compact:diquark:approx}).
}
\label{bbc:cmp:3method} }
\end{figure}

\begin{figure}[bt]
  \centerline{\epsfysize= 9.0truecm \epsfbox{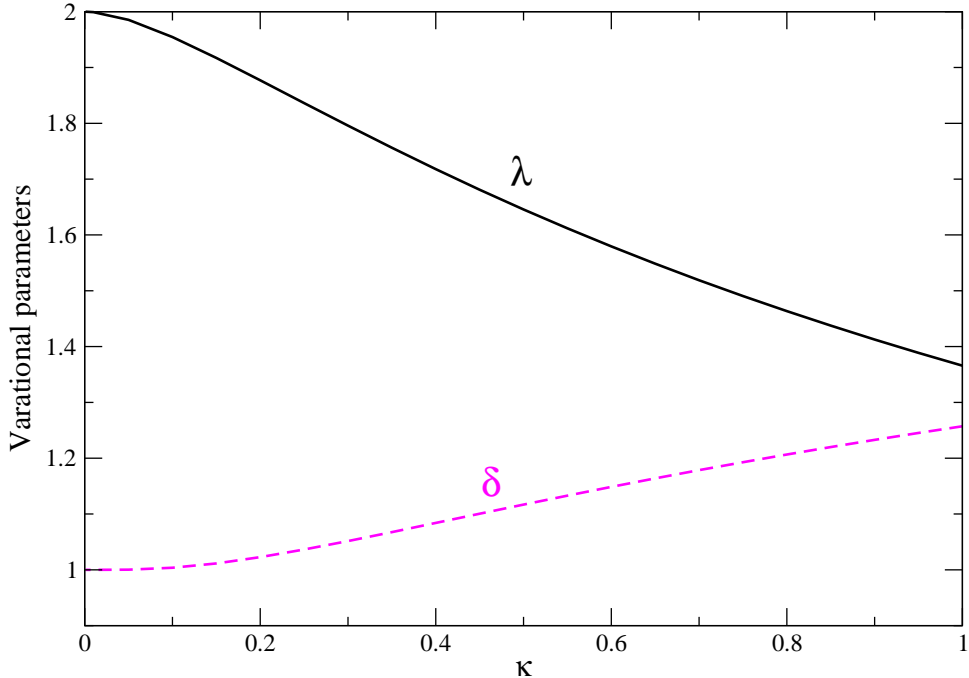}  }
 {\tighten
\caption{
Dependence of two optimized variational parameters 
$\lambda$, $\delta$ on the the mass ratio 
$m_{\rm red}/M_{\rm red}$. 
}
\label{bbc:lmdel:x} }
\end{figure}

Taking the expectation value of 
$h_S$,  (\ref{bbc:onestep:hamiton:hvby}),
in the trial state $\Psi$, 
(\ref{bbc:trial:onestep:vr:est}),
after some straightforward 
manipulation, we obtain
\bqa
E &=& -{\delta^2 \over 2} -{\kappa \lambda^2 \over 2} -
{\kappa^2 \lambda^2 \over 8}
+ \left[\, \delta (\delta-1)(1+\overline {\cal X})   \right.
\nn \\
&& \left.\,+ \kappa \left[ \lambda\,(\lambda-1) 
-\overline{\cal C}+(\lambda-2)\, 
\overline{\cal E} \right] 
+ {\kappa^2 \lambda \over 4} 
\left(\lambda + \overline{\cal E} -4 \,\delta \,\overline{\cal Y} 
\right)\right] 
/ (1+ \overline{\cal S})\,.
\label{optimum:bbc:alternative}
\eqa
The parameters in (\ref{optimum:bbc:alternative}) are given by
\bqa
   \kappa\, \overline{\cal C} &=& {\delta^3\over \pi} \int\!\! d^3 R \,
   e^{-2\delta R} \, {\cal C}(\kappa \lambda, R) 
   = {\delta \kappa\lambda\,(\delta^2+3\delta\kappa \lambda +\kappa^2\lambda^2)
   \over (\delta+\kappa\lambda)^3}  \,,
 \nn \\
   \kappa\, \overline{\cal E} &=& {\delta^3\over \pi} \int\!\! d^3 R \,
   e^{-2\delta R} \, {\cal E}(\kappa \lambda, R) 
   = { 16\,\delta^3 \kappa \lambda\,(\delta+2 \kappa\lambda) \over 
     (2 \delta+\kappa\lambda)^4 }  \,,
 \nn \\    
    \delta\, \overline{\cal X} &=& {\delta^3\over \pi} \int\!\! d^3 R \,
   e^{-2\delta R} \, {{\cal S}(\kappa \lambda, R) \over R}
   = {4\,\delta^3\,(4\delta^2+8\delta\kappa \lambda +5\kappa^2\lambda^2)
   \over (2 \delta+\kappa\lambda)^4}  \,,    
 \nn \\
   \kappa\, \overline{\cal Y} &=& {\delta^3\over \pi} \int\!\! d^3 R \,
   e^{-2\delta R} \, {\cal Y}(\kappa \lambda, R) 
   = {4\,\delta^3\,\kappa\lambda\,(2\delta+5\kappa \lambda)
   \over (2 \delta+\kappa\lambda)^5}  \,.
\label{bbc:exchange:integrals}   
 \eqa
where ${\cal C}$, ${\cal E}$ are given in (\ref{BA:C:E:integral}),
and
\bqa
 {\cal Y}(\lambda, R) &=& {\lambda^3\over \pi} \int\!\!  d^3 r\,
   e^{-\lambda (r_1+r_2)} \,  \nabla_R \,R \cdot \nabla_R\, r_1
   = {R \over 6} \,{\cal E}(\lambda, R) \,.  
\eqa

It is interesting to note that
the contribution of
the charm energy, 
which is previously computed in
Born-Oppenheimer procedure, (\ref{Eff:BO:bbc}),
is also subsumed in (\ref{optimum:bbc:alternative})
in a similar format.
Besides this, (\ref{optimum:bbc:alternative}) 
also incorporates terms that 
have been neglected in
Born-Oppenheimer approximation,
such as $\overline{\cal Y}$ 
(see the comment in Footnote~\ref{fn:BA:approx:val}).

The minimum of (\ref{optimum:bbc:alternative})
can be found by enforcing
$\partial E/\partial \lambda|_\delta
=\partial E/\partial \delta|_\lambda=0$.
It is rather difficult to derive
analytic expressions for these
optima, hence we resort to 
numerical method to determine them.
Subsequently,
the energy of $bbc$ ground state
as function of $\kappa$,
juxtaposed with predictions made by  
two other approaches,
is shown in Fig.~\ref{bbc:cmp:3method}.
The optimized values of 
$\lambda$ and  $\delta$
as functions of $\kappa$ are 
shown in Fig.~\ref{bbc:lmdel:x}.

As is expected, 
the energy predicted from this approach 
coincides with those
from the other two  
in the $\kappa \to 0$ limit.
The technical reason is easily traceable.
Note all the integrals in (\ref{optimum:bbc:alternative})
simplify greatly in this limit, {\it e.g.}, 
$\overline{\cal S},\: \overline{\cal X}\approx 1$, 
and $\overline{\cal C}, \:\overline{\cal E} \approx \lambda$.
Neglecting  higher order terms,
Eq.~(\ref{optimum:bbc:alternative}) then reduces to
\bqa
E &=& -{\delta^2 \over 2} +\delta\, (\delta-1)+\kappa \,
\left[-{\lambda^2 \over 2}+ \lambda\, (\lambda-2) \right]
+ {\cal O}(\kappa^2)\,.
\eqa
The optima $\delta=1$, $\lambda=2$ can be trivially
inferred, and the corresponding energy is exactly
the same as (\ref{compact:diquark:approx}), 
which was first derived in
the point-like diquark approximation.

Fig.~\ref{bbc:lmdel:x} illustrates some
anticipated features of a $bbc$ state.
As $\kappa$ grows, this state starts to
depart from the simple 
point-like diquark picture, 
and the effect of finite diquark size 
becomes increasingly important.
It can be clearly observed 
that $\delta$  ascends in
a slower pace 
than $\lambda$ descends.
This is compatible with the expectation
that the impact of $c$ on
$b$ is less important than 
the impact of $b$ on $c$.

One interesting observation from Fig.~\ref{bbc:cmp:3method}
is that Born-Oppenheimer approximation renders 
rather close prediction to 
that from the variational approach,
virtually in all $\kappa$ range. The reason is
perhaps that those terms dropped by Born-Oppenheimer
procedure turn out to be insignificant in this case.
In any rate, this approximation scheme
is not expected to work so well when $\kappa$ gets large.
It is worth mentioning that,
when analyzing baryon mass spectrum
from potential model approach,
Fleck and Richard have also found this scheme 
yields rather accurate results~\cite{Fleck:1989mb}.
They have attributed it to a lore that 
asserts {\it Born-Oppenheimer approximation works
always better than expected}.

We end this section by pointing out an interesting finding.
In the complicated expression for $E$,  
(\ref{optimum:bbc:alternative}), 
the last two terms 
nearly cancel with each other
in virtually all the range of $\kappa$,
once the optimized values of $\lambda$ and $\delta$ 
are used.
We thus achieve a great simplification:
\bqa
E &\approx & -{\delta^2 \over 2} - \kappa \,{\lambda^2 \over 2}\,.
\label{empirical:general:diquark:approx}
\eqa
This approximate formula works surprisingly well.
It deviates from the actual one 
by $3\, \% 
$  in maximum 
in the range $\kappa<1$. 
If one restricts to the smaller range $\kappa < 0.8$,
the error of this formula is less than
$1\, \% 
$.

Without a deeper understanding, one might 
simply regard the
success of (\ref{empirical:general:diquark:approx})
as a fortuitous coincidence. 
If taken seriously, 
it seems to indicate that
the point-like diquark picture might 
be useful
even at large $\kappa$.
It will yield the right answer,  
if one pretends that the color potential between 
two $b$ quarks is $-\delta/R$, 
and the $bb$
diquark perceived by $c$ 
is equivalent to an 
antiquark carrying the color charge $\lambda$.

\section{Phenomenology}
\label{Phenomenology}

In this section, we will assemble the knowledge 
gleaned in the preceding section to 
estimate  masses of various 
lowest-lying triply heavy baryons.
We then compare our results 
with other work in literature,
and discuss 
corresponding implications.

It should be first realized that 
our predictions will be sensitive 
to the input of heavy quark masses. 
Therefore, it is important to specify 
an appropriate quark mass scheme to 
lessen arbitrariness. Since our working assumption 
is the weak-coupling regime, it is most consistent 
to express the heavy quark pole mass in terms 
of the masses of lowest-lying quarkonia,
$J/\Psi$  and  $\Upsilon$,
assuming they are the weakly coupled system.
At order $\alpha_s^2$, 
we can write $m_c$ and $m_b$ as
\bqa
    m_c &=& {M_{J/\Psi}\over 2} 
    \left[1+ {2\, \alpha_s^2(\mu)\over 9}\right]  \,,  
\nn \\
     m_b &=& {M_{\Upsilon}\over 2} 
    \left[1+ {2\, \alpha_s^2(\mu)\over 9}\right]  \,.  
\eqa
We will take the physical values $M_{J/\psi}  = 3.097\;\:{\rm GeV}$,
$M_{\Upsilon}  = 9.460\;\:{\rm GeV}$ as input.

We are now at a position to express the 
masses of tripled-heavy baryons in 
perturbative expansion.
We start from $\Omega_{bcc}$.
Using the result of (\ref{E:bcc:variation:determ}),
we find
\bqa
     M_{\Omega_{bcc}} &=&   m_b+ 2 m_c + E
\nn \\
     &= & {M_{\Upsilon} \over 2} + M_{J/\Psi} +  
     \left[{2\over 9} M_{J/\psi} + {1\over 9} M_{\Upsilon}-
      \left({7\over 8}\right)^2\,
     {M_{J/\Psi} M_{\Upsilon} \over 2\,(M_{J/\Psi} +M_{\Upsilon})}
     \right]\, \alpha_s^2(\mu)
\nn  \\
    &= & {M_{\Upsilon} \over 2} + M_{J/\Psi} \,[1+ 0.273\, \alpha_s^2(\mu)]\,.
\label{bcc:final:mass:formula}
\eqa
In expressing the reduced mass, 
we simply replace $m_c$ with half of $M_{J/\Psi}$,
and  $m_b$ with half of  $M_\Upsilon$. 
This simplified procedure  
induces an error of order $\alpha_s^4$ to the baryon mass, 
thus legitimate at present ${\cal O}(\alpha_s^2)$ accuracy.

The masses of baryons made of three identical quarks 
can be estimated in a similar manner. 
With the input from (\ref{E:ccc:variation:determ}), 
we infer the  $\Omega_{ccc}$ mass to be
\bqa
     M_{\Omega_{ccc}} &=&   3 m_c + E
\nn \\
     &= &  
      {3 M_{J/\psi} \over 2} 
     \, \left[1+ \left({2\over9} -{0.925 \over 6} 
     \right)\alpha_s^2(\mu) \right]
\nn \\
     &= &  
      {3 M_{J/\psi} \over 2} 
     \, \left[1+ 0.068\,\alpha_s^2(\mu) \right] \,,
\label{ccc:final:mass:formula}     
\eqa
and the $\Omega_{bbb}$ mass can be obtained 
by making  obvious replacement.

For the $bbc$ state, we have attempted three different
approaches to estimate the binding energy. 
Since the one-step variational estimate
is believed to be most reliable, 
we will adopt its prediction
(though Born-Oppenheimer approximation yields 
a close result, as disclosed in Fig.~\ref{bbc:cmp:3method}).
First we need specify the value of $\kappa$, the ratio of 
two reduced quark masses. 
It can be approximated as
\bqa 
\kappa \approx {4 M_{J/\Psi}\over M_{J/\Psi}+ 2 M_\Upsilon }=0.563\,,
\eqa
and the error brought in by this procedure is 
assumed to be negligible.

The optima can be determined numerically from 
(\ref{optimum:bbc:alternative})
by variational ansatz:
\bqa
\delta &=& 1.137\,,
\hspace{0.4 in}
\lambda = 1.603\,,
\label{kappa:0.563:optima}
\eqa 
with the corresponding energy
\bqa
E &=&  -1.363 \longrightarrow -1.363 \,M_{\rm red}\,
\left({2 \alpha_s\over 3}\right)^2\,,
\eqa
where we have inserted the Bohr energy of $b$ quark 
in the last entity.

Piecing everything together, we obtain
\bqa
     M_{\Omega_{bbc}} &=&   2 m_b+  m_c + E 
\nn \\
     &= & {M_{J/\Psi} \over 2} + M_\Upsilon +  
     \left[M_{J/\psi} + (2-1.363) M_\Upsilon\right]\, 
     {\alpha_s^2(\mu) \over 9}
\nn  \\
    &= & {M_{J/\Psi} \over 2} + M_\Upsilon 
    \,[1+ 0.107\, \alpha_s^2(\mu)]\,.
\label{bbc:final:mass:formula}
\eqa

So far we have treated each of triply-heavy baryons 
separately, it is not yet clear whether there is any connection 
among them. Interestingly, there is a mass convexity 
inequality relating  different baryon states, 
which arises from 
general reasoning in QCD~\cite{Nussinov:1983hb}.
To our interest, such an inequality demands
\bqa
  M_{\Omega_{bbc}} \le 2\, M_{\Omega_{bcc}} -M_{\Omega_{ccc}}\,.
\eqa
The underlying assumption of this theorem 
is universal interquark potential. 
Taking $\alpha_s$ in (\ref{bcc:final:mass:formula}),
(\ref{ccc:final:mass:formula}) and 
(\ref{bbc:final:mass:formula}) to be equal, 
we readily verify that our predictions based on 
variational ansatz are indeed 
compatible with this QCD theorem.

There also exists another inequality,
which relates the masses of baryons 
and mesons~\cite{Nussinov:1983vh,Richard:1984wy}. 
This is derived from the assumption that 
the quark-quark potential in a baryon is a half of the
quark-antiquark potential in a meson,
which is {\it de facto} satisfied 
in Coulomb bound states.
To our purpose, this inequality reads 
\bqa
  M_{\Omega_{bbc}} \ge {M_{\Upsilon} \over 2}+  M_{B_c}\,.
\label{ineqlty:baryon:meson}  
\eqa

To make a consistent examination of 
this relation,  we need treat $B_c$ 
also as a weakly-coupled state, 
which is believed to be the case. 
Following preceding procedure, 
we can express the $B_c$ mass as
\bqa
     M_{B_c} &=&   
{M_{\Upsilon} \over 2} + {M_{J/\Psi} \over 2} +  
     \left[M_{\Upsilon}+ M_{J/\psi}  -
     {4 M_{J/\Psi} M_{\Upsilon} \over M_{J/\Psi} +M_{\Upsilon}}
     \right]\, {\alpha_s^2(\mu)\over 9}
\nn  \\
    &= & {M_{J/\Psi} \over 2}+{M_{\Upsilon} \over 2}\,
    [1+ 0.076\, \alpha_s^2(\mu)]\,.
\eqa
One can promptly check that 
this inequality also holds 
in our case.

A simple variant of (\ref{ineqlty:baryon:meson})
is to specify all the quarks to be
of a single flavor~\cite{Nussinov:1983vh}:
\bqa
  M_{\Omega_{ccc}} \ge {3 M_{J/\psi} \over 2}\,.
\label{ineqlty:ccc:Jpsi}   
\eqa
Our prediction in (\ref{ccc:final:mass:formula}) 
indeed respects this requirement.

\begin{table}[tb]
\caption{Predictions for the masses of lowest-lying triply-heavy baryons
from various work. All the masses are given in unit of $\rm GeV$. In the entries
for $\Omega_{bcc}$ and $\Omega_{bbc}$, the $J^P={1\over 2}^+$ and 
$J={3\over 2}^+$ 
partners are not distinguished since the hyperfine 
splitting has been neglected.} 
\label{tri:hvy:baryon:mass:comparison} 
\begin{center}
\begin{tabular}{ccc|ccc|ccc|ccc}
&    && & Bjorken~\cite{Bjorken:1985ei} & && This work & 
 && Vijande {\it et al}~\cite{Vijande:2004at} &
 \\ \hline 
& $\Omega_{bcc}$    &&  & $8.200\pm0.090$&  && $7.98\pm 0.07$ &  &&--&\\
& $\Omega_{ccc} $   &&  & $4.925\pm0.090$&  && $4.76\pm 0.06$ &  && 4.632 &\\
& $\Omega_{bbb}$    &&  & $14.760\pm0.180$&  && $14.37\pm 0.08$ &  && -- &\\
& $\Omega_{bbc}$    &&  & $11.480\pm0.120$  & && $11.19\pm 0.08$  &  && --&\\
\end{tabular}
\end{center}
\end{table}

To make quantitative estimates for the baryon masses,
we need specify at which scale the strong coupling
constant should be evaluated. In principle, 
physical observables should be independent of 
the choice of $\mu$, once the all-order perturbative
expansion has been worked out.
In practice, since what we have so far is 
only the leading order perturbative correction, 
our predictions are unavoidably sensitive to
the choice of $\mu$. 
To reduce the scale ambiguity optimally,
we should take $\mu$ in proximity to
the characteristic momentum transfer scale 
in a given $QQQ$ state.

It is an empirical fact that the typical momentum transfer 
scale in $J/\psi$, $B_c$
and $\Upsilon$ is about $0.9$, $1.2$ and $1.5$ GeV,
respectively.
One might expect that the corresponding scale in 
the $QQQ$ states would be considerably 
lower than that in their quarkonium counterparts. 
Encouragingly, as we have learned in 
Sec.~\ref{VC:MainBody},
the effective color strength between a pair of 
quarks gets enhanced due to the presence of the third quark.
As a result, the actual wave function is more 
compressed
than naively expected.  Therefore, 
it is not unreasonable to choose 
the scale for a $QQQ$ state close to
the one typically taken for its $Q\overline{Q}$
counterpart.
We assign $\mu=1.2$ GeV in the mass formula for 
$\Omega_{bcc}$, $\Omega_{bbb}$ and  $\Omega_{bbc}$, 
with a corresponding $\alpha_s=0.43$;
for $\Omega_{ccc}$, we take $\mu=0.9$ GeV,
with $\alpha_s=0.59$.
To compensate for our ignorance in 
uncalculated higher order corrections, 
we estimate 
the uncertainty in each mass prediction 
to be the leading ${\cal O}(\alpha_s^2)$
correction multiplied by another factor of $\alpha_s$.

Our predictions to the masses of 
various $QQQ$ ground states,
together with those made 
by other authors~\cite{Bjorken:1985ei,Vijande:2004at}, 
which employ some phenomenological 
confinement potentials,
are compiled in 
Table~\ref{tri:hvy:baryon:mass:comparison}.
The apparent discrepancy 
between the predictions 
of the  $\Omega_{ccc}$ mass
by Bjorken and by Vijande {\it et al},  
which is as large as 300 MeV, 
might reflect the large uncertainty inherent in  
phenomenological 
approaches\footnote{Note the very low $\Omega_{ccc}$ mass
predicted by Vijande {\it et al} violates 
the mass inequality (\ref{ineqlty:ccc:Jpsi}).}.
In contrast, our predictions are  
based on the perturbation theory, 
being systematically improvable, 
suffer less arbitrariness.

It can be readily recognized that Bjorken's predictions 
are systematically higher than ours.
Note the variational method by default
underestimates the binding energy, 
and a more accurate weak-coupling analysis
will predict even lower masses for 
$QQQ$ ground states,   
hence further enlarging this disagreement.

Note that the ${\cal O}(\alpha_s^2)$ corrections
in (\ref{bcc:final:mass:formula}),
(\ref{ccc:final:mass:formula}) and (\ref{bbc:final:mass:formula})
are all positive,  so as we lower down $\mu$,
which is meant to be the characteristic momentum scale, 
our predictions will shift upwards, getting close to 
Bjorken's predictions.  
When $\mu$ descends further
and becomes comparable with $\Lambda_{\rm QCD}$,
our method breaks down and 
one enters the strong-coupling
regime.  In a sense,  Bjorken's results can be 
considered as arising from a strong-coupling analysis.

The future experiments and lattice QCD simulations
will decide which prediction is closer to the reality,
consequently nature of the $QQQ$ ground states
may be disclosed.

\section{Summary and Outlook}
\label{Summary}

The theme of this work is to estimate the masses of 
various lowest-lying triply heavy baryon states, 
with the assumption
that they are weakly-coupled system, analogous to
$\Upsilon$, $B_c$ and $J/\psi$.
To achieve this, it is crucial to 
make a sound estimate for
the binding energy of a nonrelativistic three heavy quark  
system, which is bound by short-distance interquark
potentials that are organized by powers of 
$\alpha_s$ and $1/M$.  
Due to our incapability of rigorously solving 
3-body problem, we have invoked the variational method
as an approximation scheme to analyze
various $QQQ$ ground states.
As a first step, 
we have estimated the most important piece, 
{\it i.e.}, the ${\cal O}(\alpha_s^2)$ contribution 
to the binding energy, 
with only the tree-level static potential incorporated.

For the variational method to be accurate, 
it is important to choose a reasonable form of trial states.
In view of this, 
different triply-heavy baryon states, 
the $bcc$, $ccc$ ($bbb$) and $bbc$ states,
have been analyzed separately, each supplied with a 
different trial wave function motivated by
the symmetry consideration and 
the presence of hierarchy  $m_b\gg m_c$.
Inspired by the similarity between our baryonic system
and the three-body atomic system, 
some guidances have been taken from the familiar 
textbook treatment of 
helium atom and the ionized hydrogen molecule.
Among various $QQQ$ states, the $bbc$ state
is the most interesting one
but most challenging 
to analyze.
We have carried out a detailed study on this state, 
employing several different approaches. 
The implications of different approaches 
are elucidated, and in particular the 
relevance of the compact diquark picture 
has been discussed.

Masses of various $QQQ$ ground states 
derived from our formalism are compatible 
with those well-known mass inequalities in QCD.
Our quantitative predictions,
which is based on a weak-coupling treatment, 
are systematically lower than Bjorken's, 
which may instead be viewed as resulting from 
a strong-coupling analysis.
It leaves for future experiments and lattice 
QCD simulation to decide the nature of 
the lowest-lying triply heavy baryons,
whether to be weakly coupled or strongly coupled.

Besides the ability to estimate the masses, 
variational analysis also allows us to 
have a reasonable knowledge about
the quark wave functions. 
This will be useful, for example, in estimating
the hyperfine splittings in the $bcc$ and $bbc$ states. 
One important byproduct of this knowledge is that
a reasonable value of the {\it  wave function 
at the origin} can be inferred.  
Like in a heavy quarkonium, this quantity is one of 
the basic characteristics 
of a triply-heavy baryon state,
and is of phenomenological interest. 
For instance, this value is a crucial input for 
reliably estimating the fragmentation function 
for the $QQQ$ states~\cite{GomshiNobary:2004mq}.

It will be useful if alternative methods that 
have been developed to 
analyze 3-body problem, {\it e.g.}, 
the formalism of 
hyperspherical expansion~\cite{Ballot:1979hd}, 
are employed to check the accuracy 
of our results.

One apparent improvement on this work is in prospect. 
As stressed several times before, we have only incorporated
the tree-level static potential in this work, 
so that our predictions suffer from
considerable scale dependence.
The perturbative matching calculation  
for the one-loop static potential 
and tree-level spin-dependent potentials 
is straightforward.
It will  be useful to implement
their contributions into 
our variational framework.

\end{document}